\documentstyle[twocolumn,prl,aps,epsfig]{revtex}

\def\htm{\hat{t}}

\def\la{\langle}
\def\ra{\rangle}
\def\kav{\la k_T\ra}
\def\k2av{\la k_T^2\ra}
\def\knav{\la k_T^2\ra}
\newcommand{\f}[2]{\frac{#1}{#2}} 

\begin{document}

\twocolumn[\hsize\textwidth\columnwidth\hsize\csname @twocolumnfalse\endcsname
\title{Saturating Cronin effect in ultrarelativistic proton-nucleus collisions}
                                                            
\begin{flushright}
{nucl-th/9903012, KSUCNR-102-99}
\end{flushright}

\author{G\'abor Papp$^{1,2}$, P\'eter  L\'evai$^{1,3}$, and George Fai$^{1}$ }
\address{
${}^1$Center for Nuclear Research,
Department of Physics, Kent State University, Kent, Ohio 44242, USA \\ 
${}^2$HUS Research Group for Theoretical Physics, E\"{o}tv\"{o}s University, 
Budapest, P\'azm\'any P. s. 1/A, Budapest, 1117, Hungary \\
${}^3$KFKI Research Institute for Particle and Nuclear Physics, P.O. Box 49,
Budapest, 1525, Hungary }
\date{March 29, 1999}
\maketitle    

\begin{abstract}               
Pion and photon production cross sections are analyzed in proton-proton and
proton-nucleus collisions at energies 
$20$ GeV $ \lesssim \sqrt{s} \lesssim 60$ GeV.
 We separate the proton-proton and nuclear contributions to 
transverse-momentum broadening and suggest a new 
mechanism for the nuclear enhancement in the high
transverse-momentum region. 
\\
\smallskip
PACS number(s): 12.38.Bx, 13.85.Ni, 13.85.Qk, 24.85.+p, 25.75.-q 
\end{abstract}  

\vspace{0.2in}
]

\begin{narrowtext}   
\newpage 
  
Experimental results on pion and photon production in high-energy 
hadron-nucleus collisions show an extra increase 
at high transverse momentum ($p_T$) over what would be expected 
based on a simple scaling of the appropriate proton-proton ($pp$) cross
sections. The nuclear enhancement is referred to as the 
Cronin effect~\cite{Cronin}, and is most relevant at moderate
transverse momenta (3 GeV $\lesssim p_T \lesssim 6$ GeV)~\cite{Brown}. 
In relativistic nuclear collisions 
this momentum region is at the upper edge of the  $p_T$
window in Super Proton Synchrotron (SPS)
experiments at $\sqrt{s}= 20 $ AGeV,
and is measurable at 
the Relativistic Heavy Ion Collider (RHIC) at $\sqrt{s}=200$ AGeV.
The importance of a better theoretical understanding of the Cronin effect
continues to increase \cite{Wang9798,Wong98,GyulLev98} 
as new data appear from the SPS heavy-ion program
and as the commissioning of  RHIC approaches.
This calls for a systematic study of particle production moving from
$pp$ to proton-nucleus ($pA$) collisions.
In the present work we analyze the Cronin effect 
in  inclusive $\pi^0$ and $\gamma$ production.

In the past two decades the perturbative QCD (pQCD) improved
parton model has become the description of
choice for hadronic collisions at large $p_T$. 
The pQCD treatment of hadronic collisions is based on the assumption
that the composite structure of hadrons is revealed
at high energies, and the parton constituents become the
appropriate degrees of freedom for the description 
of the interaction at these energies. 
The partonic cross sections are calculable in pQCD at high energy 
to leading order (LO) or next-to-leading-order (NLO) 
\cite{Field,NLOAur,GorVogel94}.
The parton distribution function (PDF) 
and the parton fragmentation function (FF), however,  
require the knowledge of non-perturbative QCD and are not
calculable directly by present techniques. The PDFs and FFs, which are
believed to be universal, are fitted to reproduce the data
obtained in different reactions.
In recent NLO calculations the various scales 
($Q$,$\Lambda_{QCD}$, etc.) are optimized 
to improve the agreement between data and 
theory~\cite{OptimQres,Aur98}.

In theoretical investigations of $\pi^0$ and $\gamma$ 
production in $pA$
collisions another method appeared and became
popular~\cite{Huston95,E706,Ziel98}: the different scales of the pQCD
calculations are fixed and the NLO pQCD theory is supplemented by an
additional non-perturbative parameter, the {\it intrinsic transverse
momentum} ($k_T$) of the partons. 
The presence of an intrinsic transverse momentum, 
as a Gaussian type broadening of the
transverse momentum distribution of the initial state partons in colliding
hadrons was investigated as soon as pQCD calculations were applied to
reproduce large-$p_T$ hadron production~\cite{ktold,Owens87}. The
average intrinsic transverse momentum needed was small, 
$\kav\sim0.3-0.4$ GeV, and could be easily understood in terms of
the Heisenberg uncertainty relation for partons inside the proton. 
This simple physical
interpretation was ruled out as the only source of intrinsic $k_T$
by the analysis of new experiments on direct photon production, where 
$\kav\sim1$ GeV was obtained in the fix target 
Tevatron experiments~\cite{Huston95,E706,Ziel98} and $\kav\sim 4$ GeV 
was found at the Tevatron collider for muon, photon and jet
production~\cite{Ziel98}. 
New theoretical efforts were ignited to
understand the physical origin of $\kav$~\cite{Guo96,Lai98}.
Parallel to these developments, $k_T$ smearing was applied successfully to
describe 
ultrarelativistic nucleus-nucleus collisions~\cite{Wang9798,Wong98} and
$J/\psi$ production at Tevatron and HERA~\cite{Srid98}.
One possible explanation of the enhanced
$k_T$-broadening is in terms of multiple gluon
radiation \cite{Lai98}, which makes $\kav$ 
reaction and energy dependent.
In the absence of a full theoretical description,
intrinsic $k_T$ can be used phenomenologically in $pp$ collisions.
A reasonable reproduction of the $pp$ data is a prerequisite for the
isolation of the nuclear enhancement we intend to focus on.

In the lowest-order pQCD parton model,
direct pion production can be described in $pp$ collisions by
\begin{eqnarray}
\label{fullpi}
  E_{\pi}\f{d\sigma_\pi^{pp}}{d^3p} &=&
        \sum_{abcd}\!  \int\!dx_1 dx_2\ f_{a/p}(x_1,Q^2)\
        f_{b/p}(x_2,Q^2)\ \nonumber \\
          && \ \ \ \  K  \f{d\sigma}{d\htm}(ab\to c d)\,
   \frac{D_{\pi/c}(z_c,{\widehat Q}^2)}{\pi z_c} \ \ \  , 
  \end{eqnarray}
where  $f_{a/p}(x,Q^2)$ and  $f_{b/p}(x,Q^2)$  are the PDFs for the
colliding partons $a$ and $b$ in the interacting protons 
as functions of momentum fraction $x$ and momentum transfer $Q$, and
$\sigma$ is the LO hard scattering cross section of the appropriate
partonic subprocess.
The K-factor accounts for higher order corrections~\cite{Owens87}. 
Comparing LO and NLO calculations one can
obtain a constant value, $K\approx 2$, 
as a good approximation of the higher order contributions 
in the $p_T$ region of interest~\cite{Wong98,EskW95}.
In eq.(\ref{fullpi}) 
$D_{\pi/c}(z_c,{\widehat Q}^2)$ is the FF for the pion,
with the scale ${\widehat Q} = p_T /z_c$, where 
$z_c$ indicates the momentum fraction of the final hadron. 
We use a NLO parameterization of the FFs~\cite{BKK}.
Direct $\gamma$ production is described similarly~\cite{Field}.

The generalization to incorporate the $k_T$ degree of freedom is
straightforward~\cite{Wang9798,Wong98}. Each integral over the parton
distribution functions is extended to $k_T$-space,
\begin{equation}
\label{ktbroad}
dx \ f_{a/p}(x,Q^2) \rightarrow dx 
\ d^2\!k_T\ g({\vec k}_T) \  f_{a/p}(x,Q^2) \ ,
\end{equation}
and, as an approximation,
$g({\vec k}_T)$ is taken to be a Gaussian:
$g({\vec k}_T) = \exp(-k_T^2/\langle k_T^2 \rangle)/
{\pi \langle k_T^2 \rangle}$.
Here $\langle k_T^2 \rangle$ is the 2-dimensional width of the $k_T$
distribution and it is related to the average transverse momentum of one parton
as $\langle k_T^2 \rangle = 4 \langle k_T \rangle^2 /\pi$.

\begin{figure}
\vskip -0.4in
\epsfxsize=3.5in
\epsfysize=3.3in
\centerline{\epsffile{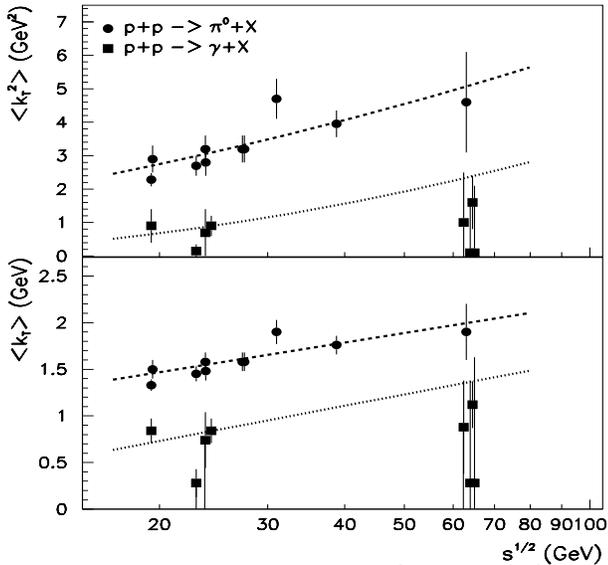}}
\vskip -0.05in
\caption[]{
 \label{figure1}
 The best fit values of $\knav$ in $pp \rightarrow \pi^0 X$ 
\cite{Cronin,E704pi,NA24pi,R110pi,E605pi,WA70,R807}
and $p p \rightarrow \gamma X$ reactions
\cite{WA70,R807,NA24,UA6,R108,R110,R806}
(upper panel), and a calculated $\kav$
(lower panel).
See text for the lines.
}
\end{figure}

We applied this model to describe the measured data
in $pp\rightarrow \pi^0  X$ reactions 
\cite{Cronin,E704pi,NA24pi,R110pi,E605pi,WA70,R807}.
If $\pi^+$ and $\pi^-$ production was measured, 
we constructed  the combination
$\pi^0=(\pi^+ + \pi^-)/2$, given by the FFs \cite{BKK} we use.
The calculations were corrected for the finite rapidity windows of 
the data.
The Monte-Carlo integrals were carried out by the
standard VEGAS-routine~\cite{VEGAS}.
For the PDFs we used the MRST98 set\cite{MRST98}, which
incorporates an intrinsic $k_T$.
The scales are fixed,
$\Lambda_{\overline {MS}}(n_f=4) = 300$ MeV and $Q=p_T/2$.
We fitted the data minimizing  $\Delta^2=\sum (Data-Theory)^2/Theory^2$
in the midpoints of the data. Fig.1. shows the obtained fit values for
$\knav$.   The error bars display a 
$\Delta^2= \Delta^2_{min}\pm 0.1$ uncertainty in the fit procedure.
The uncertainty is small at $\sqrt{s}=20-30$ GeV and relatively large
at $\sqrt{s} \approx 60$ GeV, indicating that
$\knav$ is more sharply determined at lower energies.
We use a value 
of \mbox{$\knav = 3$  GeV$^2$} for $\pi^0$ at $\sqrt{s}=27.4$ GeV. 
The value of $\knav$ appears to increase with energy. 
The dashed line serves to guide the eye and indicates a linear
increase  to a value of $\kav= 3.5$ GeV at $\sqrt{s}=1800$ GeV
\cite{Ziel98}.
  
Furthermore, we analyzed the data  from
 $pp\rightarrow \gamma X$ reactions
\cite{WA70,R807,NA24,UA6,R108,R110,R806}.
These results are also shown in Fig.1. 
In this case much lower values are obtained for $\knav$
with large uncertainty. 
The dotted line represents the results obtained in Ref.~\cite{Ziel98}
from diphoton and dimuon data.
In the following  we  use $\knav=$1.2 and 1.5 GeV$^2$  
at $\sqrt{s}=31.6$  and $ 38.8$ GeV, respectively.

Fig.2. displays the ratio of data to our calculations
as a function of $x_T=2p_T/\sqrt{s}$
for $pp \rightarrow \pi^0 X$,
applying the obtained best fit values for $\knav$ in 
the different experiments.
For $pp \rightarrow \gamma X$ data we achieve a similarly good
agreement.
Further details of the comparisons with data and with other calculations 
will be published elsewhere \cite{PLF99}.

\begin{figure}
\vskip -0.4in
\epsfxsize=3.5in
\epsfysize=3.3in
\centerline{\epsffile{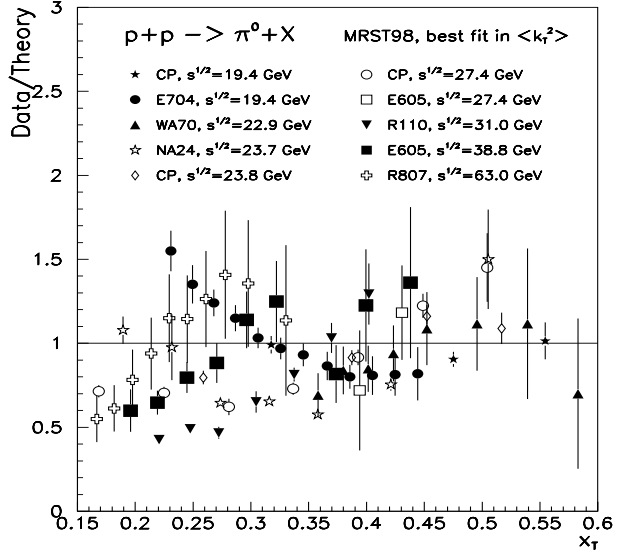}}
\vskip -0.05in
\caption[]{
 \label{figure2}
 The ratio of data to theory 
for $p p \rightarrow \pi^0 X$ reactions
\cite{Cronin,E704pi,NA24pi,R110pi,E605pi,WA70,R807}
applying the best fit for $\knav$, $x_T=2p_T/\sqrt{s}$.
}
\end{figure}

With $\pi^0$ and $\gamma$ production in $pp$ collision reasonably
under control, we turn to the nuclear enhancement in $pA$ collisions.
In minimum-biased $pA$ collisions the pQCD description 
of the inclusive pion cross section is based on
\begin{eqnarray}
\label{fullpipA}
 && E_{\pi}\f{d\sigma_\pi^{pA}}{d^3p} =\!
        \sum_{abcd}\!  \int\!\! d^2\!b\, t_A(b)\!
        \int\!\!dx_1 dx_2 d^2k_{T,a}d^2k_{T,b}\ 
        g(\vec{k}_{T,a}) 
	\nonumber \\
        && \ \ g(\vec{k}_{T,b}) f_{a/p}(x_1,Q^2) f_{b/p}(x_2,Q^2)\ 
           K  \f{d\sigma}{d\htm}
   \frac{D_{\pi/c}(z_c,{\widehat Q}^2)}{\pi z_c} \,.
\end{eqnarray}
Here $b$ is the impact parameter and
$t_A(b)$ is the nuclear thickness function normalized as
$\int d^2 b \ t_A(b) = A$. For simplicity, we  use 
a sharp sphere nucleus with 
$t_A(b) = 2 \rho_0 \sqrt{R_A^2-b^2} $, where $R_A=1.14 A^{1/3}$
and $\rho_0=0.16$ fm$^{-3}$. 
Next we discuss the nuclear enhancement of $\knav$.

The standard physical explanation of the Cronin effect \cite{Wang9798}
is that the proton traveling
through the nucleus gains extra transverse momentum due to 
random soft collisions and the partons enter the final hard
process with this extra  $k_T$. 
In our approximation initial 
soft processes increase the value of $\knav$, but
this effect does not depend on the scale, $Q^2$, of the
hard process to occur later.
However, it may depend on
the initial center-of-mass energy, $\sqrt{s}$. 
Furthermore, it is important to note, that  in our description
not all participant protons are 
automatically endowed with the extra $\knav$ enhancement,
only the parton distribution of the colliding protons is
affected according to the number of soft 
collisions suffered.
To characterize the $\knav$ enhancement, we write  the
width of the transverse momentum distribution of the partons
in the incoming proton as
\begin{equation}
\label{ktbroadpA}
\knav_{pA} = \knav_{pp} + C \cdot h_{pA}(b) \ . 
\end{equation}
Here $h_{pA}(b)$ describes the number of {\it effective} 
nucleon-nucleon (NN) collisions at impact parameter $b$
which impart an average transverse momentum squared $C$. 

Naively  all possible soft interactions are included,
$h^{all}_{pA}(b)=\nu_A(b) - 1$,
where $\nu_A(b) = \sigma_{NN} \ t_A(b)$ 
is the collision number at impact parameter $b$ with $\sigma_{NN}$
being the total inelastic cross section.
Applying this model to $\pi^0$ production 
in $pA$ ($A=Be,Ti,W$) collisions at 
$\sqrt{s}=27.4$ GeV \cite{Cronin},
we extract $C^{all}_{pBe} = $ \mbox{$0.8\pm 0.2$ GeV$^2$}, 
$C^{all}_{pTi} = 0.4 \pm 0.2$ GeV$^2$, and 
$C^{all}_{pW} = 0.3 \pm 0.2$  GeV$^2$.
The target dependence of the extra enhancement in $\knav$ 
is inconsistent with the assumption of a target-independent average 
transverse momentum transfer per NN collision and/or with the 
form of $h^{all}_{pA}(b)$.

Inspired by the $A$-dependence, we propose another
physical picture of the nuclear enhancement effect. 
According to this mechanism, the incoming nucleon first
participates in a semi-hard ($Q^2 \sim 1$ GeV$^2$) collision 
resulting in an increase of the width of its $k_T$ distribution.
There is at most one collision able to impart this increased
value of $\knav$, characteristic of the no-longer coherent
nucleon. Further soft or semi-hard collisions do not affect the
$k_T$ distribution of the partons in the incoming nucleon.
This saturation effect can be approximated well by a smoothed step
function $h^{sat}_{pA}(b)$ defined as

\[ h^{sat}_{pA}(b)  = \left\{ \begin{array}{ll}
           0  & \mbox{ if \ $\nu_A(b) < 1$ } \\
  \nu_A(b)-1  & \mbox{ if \ $1 \leq \nu_A(b) < 2$ } \ \ \ . \\
           1  & \mbox{ if \ $2 \leq \nu_A(b)$  } 
           \end{array} \right. \]
The saturated Cronin factor is denoted by  $ C^{sat}$.

Using the same $pA$ data as previously \cite{Cronin},
$C^{sat}=1.2$ GeV$^2$ gives a good fit for all three
targets, $Be$, $Ti$ and $W$. Fig.3. displays 
our results with (full line) and without (dashed line) the 
Cronin enhancement.  
The lower panel shows the data/theory ratio on a linear scale for the
$pW$ case. It is interesting to note that with the saturated Cronin effect
the remaining deviation from one in the data/theory ratio is 
similar in shape 
to the $p_T$-dependent K factor obtained in Ref.~\cite{Wong98}.
The $pBe\rightarrow \pi^0 X$ data at energies 
$\sqrt{s}=31.6$ GeV and 38.8 GeV \cite{E706}
are also described quite well with the same saturating Cronin effect
\cite{PLF99}.

\begin{figure}
\epsfxsize=3.4in
\epsfysize=3.3in
\centerline{\epsffile{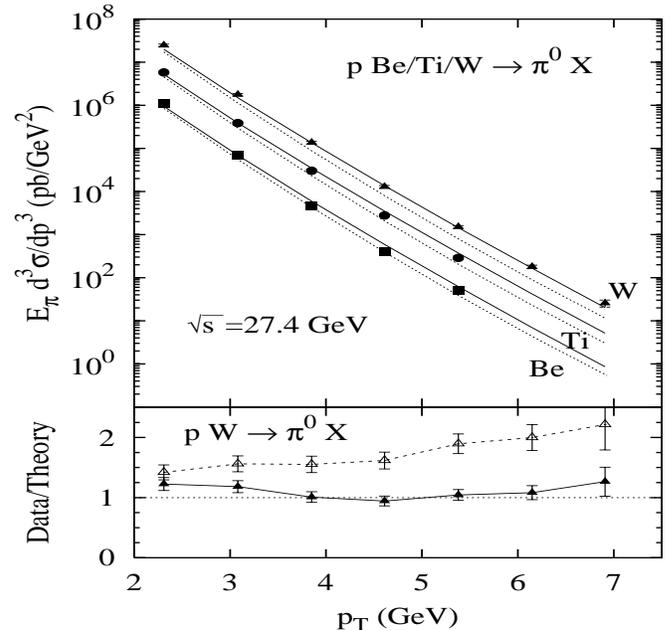}}
\caption[]{
 \label{figure3}
Cross section per nucleon in 
 $pA \rightarrow \pi^0 X$ reactions ($A=Be,Ti,W$), data from
\cite{Cronin}. We show 
$C^{sat}=1.2$ GeV$^2$ (full lines) and
$C^{sat}=0$ (dashed lines). The lower panel displays the
data to theory ratio on a linear scale for the $pW$ collision.
}
\end{figure}

Let us now discuss  photon production in the same energy region.
We speculate that the
nuclear enhancement does not depend on the 
outgoing particle, and thus  use the same  
$C^{sat}$  for $\gamma$ as for $\pi^0$ production.
Fig.4. shows the $pBe \rightarrow \gamma X$ reaction  
at energies $\sqrt{s}=31.6$ GeV and 
$\sqrt{s}=38.8$ GeV. The data are from Ref.~\cite{E706}.
The calculations are carried out with 
$C^{sat}=1.2$ GeV$^2$ (full lines). For comparison we show the results
without nuclear enhancement, $C^{sat}=$0 (dashed lines).
In the lower panel the data/theory ratio is displayed for
$\sqrt{s}=31.6$ GeV.

The common value of $C^{sat}$  in the studied energy range
indicates that the extra enhancement
in $pA$ collision is independent of the produced final state particle.
We interpret this $C^{sat}$ as the 
square of the characteristic transverse
momentum imparted in one semi-hard collision prior to the hard
scattering.
Independent of the details of the mechanism,
the extra nuclear enhancement in eq.~(4) appears to have the
same total value, which 
is on the scale of the intuitive dividing line between hard
and soft physics, $C\cdot h_{pA}(0) \approx 1$ GeV$^2$.
For RHIC predictions it would be necessary to see the energy
dependence of $C^{sat}$, and whether the same nuclear enhancement 
is obtained at much higher energies.

\begin{figure}
\epsfxsize=3.4in
\epsfysize=3.3in
\centerline{\epsffile{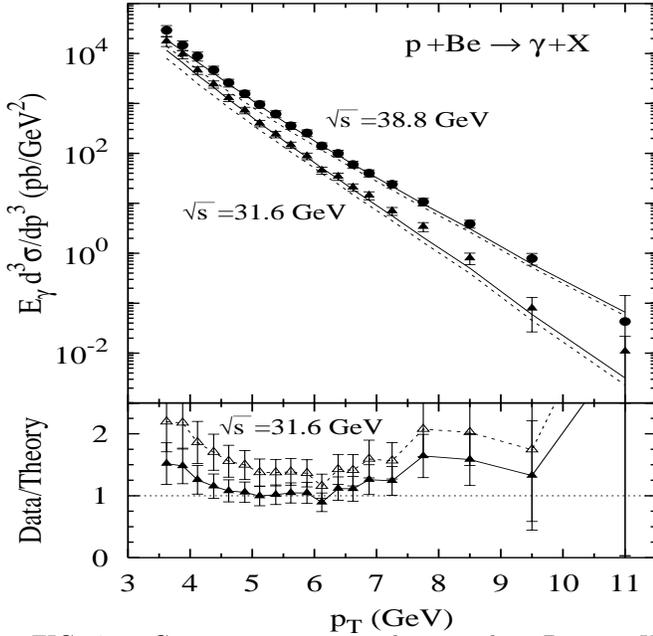}}
\caption[]{
 \label{figure4}
Cross section per nucleon in the 
 $pBe \rightarrow \gamma X$ reaction 
at $\sqrt{s}=38.8$ GeV (dots)
and at $\sqrt{s}=31.6$ GeV (full triangles) \cite{E706}.
Solid lines indicate $C^{sat}=1.2$ GeV$^2$, dashed lines mean $C^{sat}=0$.
Lower panel shows data/theory for $\sqrt{s}=31.6$ GeV.
}
\end{figure}

In the present letter we separated the $pp$ and nuclear
contributions to the width of the parton transverse momentum distribution,
and proposed a saturating model of the Cronin effect in
proton-nucleus collisions. According to our picture, 
the incoming proton suffers at most one semi-hard
scattering prior to the hard parton scattering.
In the semi-hard  collision the width of the 
transverse momentum distribution of the partons inside the incoming proton
increases.
This prescription
describes the $p_T$ dependence of the nuclear enhancement in
$\pi^0$ and $\gamma$ production remarkably well.
We are confident that complete NLO calculations, which are
left for future work, will improve
the agreement further.
Systematic $pA$ experiments are needed 
to determine the energy dependence of $C^{sat}(s)$
and to extrapolate to $AA$ collisions at RHIC.

We thank X.N. Wang and C.Y. Wong for stimulating discussions
and M. Begel and the E706 Collaboration for providing their data.
We are grateful to G. David for his continued interest in the project. 
Work supported in part by DOE grant
No. DE-FG02-86ER40251,  by US/Hungarian Science and 
Technology Joint Fund No.652/1998,
and OTKA Grant No. F019689.

\end{narrowtext} 
\end{document}